\documentclass[useAMS]{mn2e}

\usepackage{latexsym,graphicx}%,natbib}
\usepackage[euler]{textgreek}
\usepackage{textgreek}
\usepackage{color}

\newcommand\kms{\rm\,km\,s^{-1}}
\newcommand\msun{\rm\,M\sun}
\newcommand\lsun{\rm\,L\sun}
\newcommand\rsun{\rm\,R\sun}
\newcommand\myr{\msun \, {\rm yr}^{-1}}
\newcommand{\MC}{\multicolumn}

\def\apgt{\ {\raise-.5ex\hbox{$\buildrel>\over\sim$}}\ }
\def\aplt{\ {\raise-.5ex\hbox{$\buildrel<\over\sim$}}\ }

\title[CPD$-$64\degr2731: a massive rejuvenated runaway star]{CPD$-$64\degr2731: a massive spun-up and rejuvenated high-velocity
runaway star}\author[V. V.~Gvaramadze et al.]
        {V. V.~Gvaramadze,$^{1,2,3}$\thanks{E-mail: vgvaram@mx.iki.rssi.ru} O. V.~Maryeva,$^{4,1}$ A. Y.~Kniazev,$^{5,6,1}$ 
        D. B.~Alexashov,$^{7,2}$ \\ 
        \newauthor N.~Castro,$^{8}$ N.~Langer$^{9}$ and I. Y.~Katkov$^1$\\
        $^1$Sternberg Astronomical Institute, Lomonosov Moscow State University, Universitetskij Pr. 13, Moscow 119992, Russia\\
        $^2$Space Research Institute, Russian Academy of Sciences, Profsoyuznaya 84/32, 117997 Moscow, Russia \\
        $^3$Isaac Newton Institute of Chile, Moscow Branch, Universitetskij Pr. 13, Moscow 119992, Russia \\
        $^4$Astronomical Institute, Czech Academy of Sciences, Fri\v{c}ova 298, 251 65 Ond\v{r}ejov, Czech Republic \\ 
        $^5$South African Astronomical Observatory, PO Box 9, 7935 Observatory, Cape Town, South Africa \\
        $^6$Southern African Large Telescope Foundation, PO Box 9, 7935 Observatory, Cape Town, South Africa \\
        $^7$Institute for Problems in Mechanics, prosp. Vernadskogo 101, block 1, Moscow 119526, Russia\\
        $^8$Department of Astronomy, University of Michigan, 1085 S. University Avenue, Ann Arbor, MI 48109-1107, USA \\
        $^9$Argelander-Institut f\"ur Astronomie, Auf dem H\"ugel 71, D-53121 Bonn, Germany
        }
\begin{document}

\date{Accepted 2018 October 26. Received 2018 October 12; in original form 2018 September 14}

\maketitle

\label{firstpage}

\begin{abstract}
We report the results of our study of the high-velocity ($\approx160 \, \kms$) runaway O star CPD$-$64\degr2731 and its associated 
horseshoe-shaped nebula discovered with the {\it Wide-field Infrared Survey Explorer}. Spectroscopic observations with the 
Southern African Large Telescope and spectral analysis indicate that CPD$-$64\degr2731 is a fast-rotating main-sequence 
O5.5 star with enhanced surface nitrogen abundance. We derive a projected rotational velocity of 
$\approx300\, \kms$ which is extremely high for this spectral type. Its kinematic age of $\approx6$\,Myr, assuming it was born near 
the Galactic plane, exceeds its age derived from single star models by a factor of two. These properties suggest that CPD$-$64\degr2731 
is a rejuvenated and spun-up binary product. The geometry of the nebula and the almost central location of the star within it argue 
against a pure bow shock interpretation for the nebula. Instead, we suggest that the binary interaction happened recently, thereby 
creating the nebula, with a cavity blown by the current fast stellar wind. This inference is supported by our results of 2D numerical 
hydrodynamic modelling.
\end{abstract}


%---------------------------------------------------------------------
\section{Introduction}
%---------------------------------------------------------------------
\label{sec:int}

%%%%%%%%%%%%%%%%%%%%%%%%%%%%%%%%%%%%%%%%%%%%%%%%%%%%%%%%%%%%%%%%%%%%%%%%%%%
\begin{figure*}
\includegraphics[width=14cm]{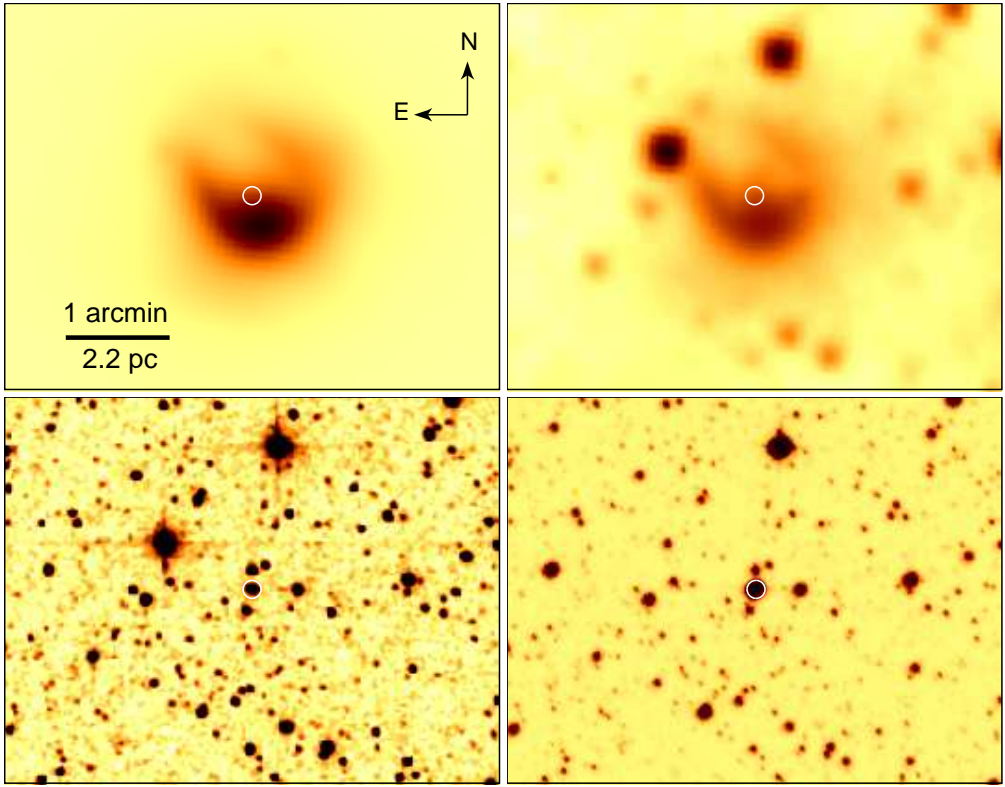}
\centering \caption{From left to right, and from top to bottom: {\it WISE} 22 and 12\,\micron, 2MASS $K_{\rm s}$-band and 
SHS H$\alpha$ images of the region containing CPD$-$64\degr2731 (marked by a circle) and its surrounding infrared nebula. 
The orientation and the scale of the images are the same.} 
\label{fig:neb}
\end{figure*}
%%%%%%%%%%%%%%%%%%%%%%%%%%%%%%%%%%%%%%%%%%%%%%%%%%%%%%%%%%%%%%%%%%%%%%%%%%%

Wind parameters of (single) massive stars, such as the terminal wind velocity and mass-loss rate, undergo significant changes
in the course of stellar evolution, which can lead to the formation of transient circumstellar nebulae around evolved massive 
stars. For example, the origin of circumstellar nebulae around late nitrogen-sequence Wolf-Rayet (WNL) stars could be attributed 
to the interaction between the fast wind of these stars and a slow dense material shed during preceding evolutionary stages 
(e.g. Garc\'ia-Segura, Mac Low \& Langer 1996a; Garc\'ia-Segura, Langer \& Mac Low 1996b; Brighenti \& D'Ercole 1997). The 
circumstellar nebulae are generally observed around field (i.e. runaway) stars or around stars at the outskirts of their 
parent star clusters, where winds from other cluster members do not prevent their formation.

Observations show that the majority of massive stars are formed in binary and multiple systems (Mason et al. 2009; Sana \& 
Evans 2011; Chini et al. 2012). The mass loss from these systems is defined not only by the (variable) stellar wind of 
individual companion stars, but also by various binary interaction processes, such as mass transfer or merger. 
In both cases, the system can lose a significant fraction of its mass (up to several solar masses) which slowly
spreads into the circumstellar environment (e.g. Petrovic, Langer \& van der Hucht 2005; Glebbeek et al. 2013). After the 
thermal adjustment on the Kelvin-Helmholtz time-scale, the mass gainer or merger product might become a source of fast wind,
which will blow a bubble in the surrounding medium. This bubble would be surrounded by a shell (nebula) provided that its
expansion is supersonic. The shell will not form if the binary interaction occurred within a larger hot bubble produced by 
winds of all massive stars of the parent star cluster. Thus, to produce a detectable nebula, the binary system should find 
itself far from the birth cluster well prior to the beginning of the mass transfer or merger event. 

Observations also show that $\approx70$ per cent of runaway O stars are binary systems (Chini et al. 2012). Moreover, about 
40 per cent of O stars interact with a companion while they are still on the main sequence (Sana et al. 2012). This implies 
that many of the main-sequence O stars in the field could be the rejuvenated products of mass transfer or binary 
mergers (runaway blue stragglers), and therefore some of them could be surrounded by initially almost spherical nebulae, whose 
origin hardly can be explained within the framework of single star evolution. These nebulae, however, should be 
extremely rare because they rapidly transform into bow shocks when the forward edge of the wind bubble crosses the region 
occupied by the slow material lost during the binary interaction process and starts to interact with the unperturbed 
interstellar medium (ISM).

In this paper, we report the discovery of an almost complete mid-infrared shell around the high-velocity runaway main-sequence 
O star CPD$-$64\degr2731. In Section\,\ref{sec:neb}, we present images of the nebula and review observational data on the star. 
Section\,\ref{sec:obs} describes our optical spectroscopic observations of CPD$-$64\degr2731. The spectral analysis of the 
star is given in Section\,\ref{sec:ana}. In Section\,\ref{sec:run}, we show that CPD$-$64\degr2731 is a high-velocity runaway 
star, while in Section\,\ref{sec:blu}, we argue that CPD$-$64\degr2731 is a spun-up and rejuvenated product of binary interaction 
(massive blue straggler). In Sections\,\ref{sec:ori} and \ref{sec:num}, we discuss possible explanations for the origin of 
the nebula and verify them by means of 2D numerical simulations, respectively. Concluding remarks are given in 
Section\,\ref{sec:con}. We summarize in Section\,\ref{sec:sum}. 

\section{Horseshoe-shaped nebula and its central star}
\label{sec:neb}

The nebula, which is the subject of this paper, was discovered in the Circinus constellation during our search for mid-infrared 
nebulae in areas of the sky that were not covered by the major {\it Spitzer Space Telescope} surveys of massive star-forming 
regions (for motivation of the search see Gvaramadze, Kniazev \& Fabrika 2010). For this purpose, we used data from the all-sky 
survey carried out by the {\it Wide-field Infrared Survey Explorer} ({\it WISE}), which provides images at four wavelengths: 
3.4, 4.6, 12 and 22\,\micron, with angular resolution of 6.1, 6.4, 6.5 and 12.0 arcsec, respectively (Wright et al. 2010). 

The search has led to the discovery of many dozens of previously unknown nebulae. Some of them were found around newly 
identified massive stars, such as luminous blue variables and blue supergiants (e.g. Gvaramadze et al. 2012a). Some others 
were interpreted as bow shocks produced by known or newly identified OB stars (e.g. Gvaramadze et al. 2011). As a by-product, 
we discovered three planetary nebulae (e.g. Gvaramadze \& Kniazev 2017), one of which, indicated in the SIMBAD data 
base\footnote{http://simbad.harvard.edu/simbad/} as a possible planetary nebula ``NAME PN\,Ra\,7'' (Ferrero et al. 2015), 
was found to be associated with a [WO1] star (Gvaramadze et al., in preparation). And one of the nebulae detected with 
{\it WISE} turns out to be quite unusual because it is associated with a high-velocity runaway main-sequence
O star, but did not look like a genuine bow shock.

Figure\,\ref{fig:neb} shows {\it WISE} 22 and 12\,\micron \, images of this nebula. In both images it appears as a 
horseshoe-shaped shell with angular radius of $\approx30$ arcsec and enhanced brightness along the southern rim. The 
shell is not visible at shorter {\it WISE} wavelengths, nor in the optical images from the Digitized Sky Survey\,II 
(McLean et al. 2000) and the SuperCOSMOS H-alpha Survey (SHS; Parker et al. 2005). The horseshoe appearance of the 
shell and its brightness asymmetry suggest that the star producing the shell is moving to the south or southwest, and 
that the geometry of the shell is affected by the ram pressure of the incoming ISM flow. 

Using the SIMBAD data base, we identified the nebula with the {\it Infrared Astronomical Satellite} ({\it IRAS}) source 
IRAS\,14032$-$6515, and found within it a massive star, known as CPD$-$64\degr2731 or ALS\,3198 (marked in 
Fig.\,\ref{fig:neb} by a circle). CPD$-$64\degr2731 was recognized as an OB star by Lynga (1969) and later on was 
classified as a fast-rotating O5 giant (O5\,IIIn) star by Vijapurkar \& Drilling (1993). In Fig.\,\ref{fig:neb}, we also 
show the $K_{\rm s}$-band and H$\alpha$ images of the field containing the nebula and CPD$-$64\degr2731 taken, respectively, 
from the Two Micron All Sky Survey (2MASS; Skrutskie et al. 2006) and the SHS. One can see that the star is slightly offset 
(by $\approx5$ arcsec) from the geometric centre of the nebula towards its brighter side. 

The details of CPD$-$64\degr2731 are summarized in Table\,\ref{tab:det}.

%%%%%%%%%%%%%%%%%%%%%%%%%%%%%%%%%%%%%%%%%%%%%%%%%%%%%%%%%%%%%%%%%%%%%%%%%%%
\begin{table}
  \caption{Details of CPD$-$64\degr2731. The spectral type, SpT, is based on our
spectroscopic observations and spectral analysis. The coordinates and the $JHK_{\rm s}$ photometry are from 
the 2MASS All-Sky Catalog of Point Sources (Cutri et al. 2003), the $BV$ photometry is from the AAVSO Photometric 
All-Sky Survey (Henden et al. 2016), and the [3.4] and [4.6] magnitudes are from the AllWISE Data Release (Cutri et 
al. 2014).}
  \label{tab:det}
  \begin{center}
  \begin{tabular}{lc}
      \hline
      SpT & O5.5\,Vn((f)) \\
      $\alpha$ (J2000) & $14^{\rm h} 07^{\rm m} 07\fs52$ \\
      $\delta$ (J2000) & $-65\degr 29\arcmin 34\farcs4$ \\
      $l$ & 310\fdg6792\\
      $b$ & $-$3\fdg7659 \\
      $B$ (mag) & $10.872\pm0.020$ \\
      $V$ (mag) & $10.597\pm0.015$ \\
      $J$ (mag) & $10.053\pm0.024$ \\
      $H$ (mag) & $9.998\pm0.025$ \\
      $K_{\rm s}$ (mag) & $9.990\pm0.023$ \\
      $[3.4]$ (mag) & $9.926\pm0.024$ \\
      $[4.6]$ (mag) & $9.965\pm0.022$ \\
      \hline
    \end{tabular}
\end{center}
\end{table}
%%%%%%%%%%%%%%%%%%%%%%%%%%%%%%%%%%%%%%%%%%%%%%%%%%%%%%%%%%%%%%%%%%%%%%%%%%%

\section{Spectroscopic observations}
\label{sec:obs}

CPD$-$64\degr2731 was observed with the Southern African Large Telescope (SALT) High Resolution Spectrograph (HRS; Barnes 
et al. 2008; Bramall et al. 2010, 2012; Crause et al. 2014) at Medium Resolution (MR) mode on 2018 February 1 to verify its 
spectral type, and re-observed on 2018 February 25 and July 18 (during twilight time) to search for possible radial velocity 
variability. All observations were carried out with the same single exposure of 1200\,s. The seeing during these observations 
was, respectively, $\approx1.6$, 2.6 and 1.6 arcsec. 

\begin{figure*}
\begin{center}
\includegraphics[width=12cm,angle=270,clip=]{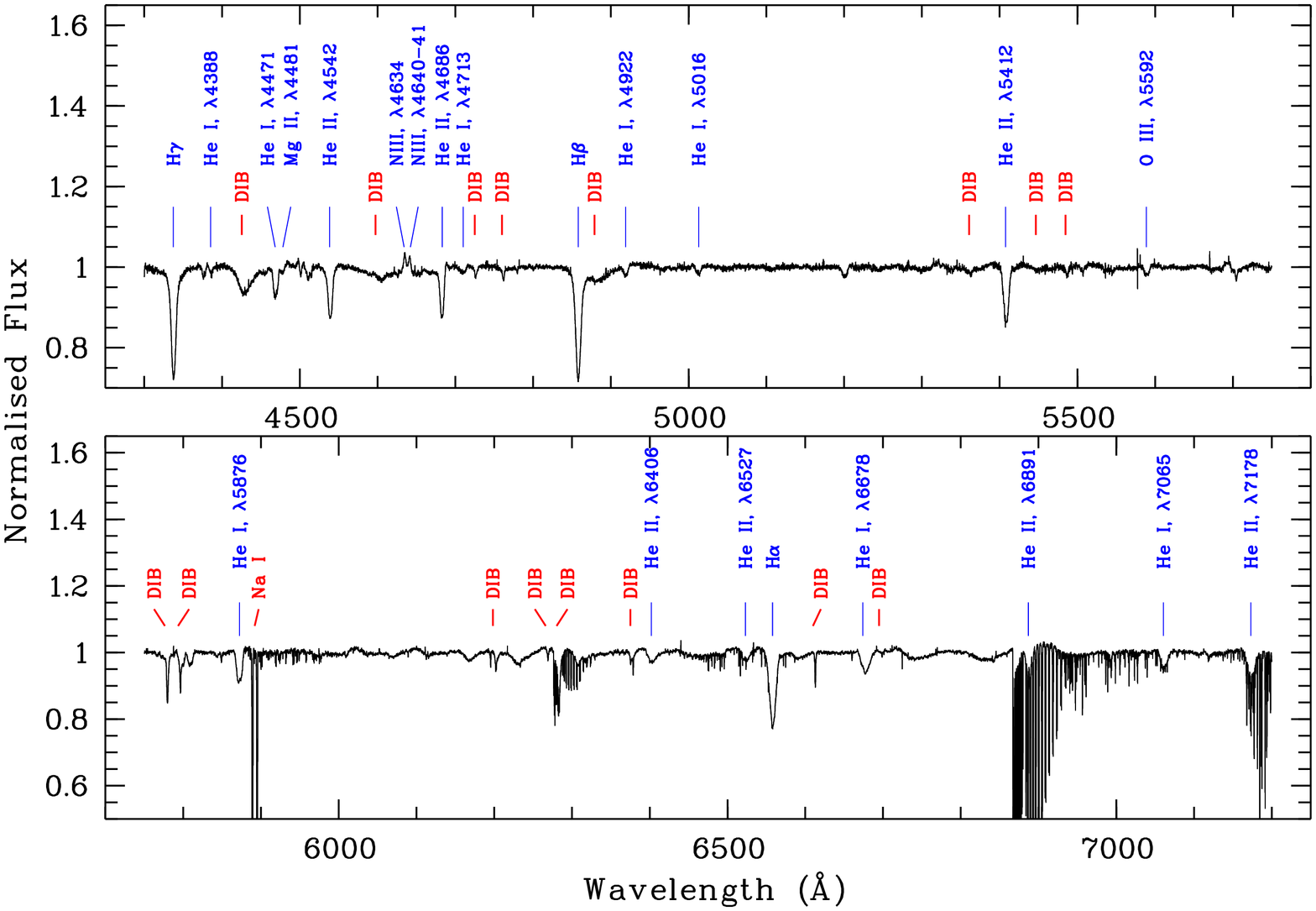}
\end{center}
\caption{A part of the normalized spectrum of CPD$-$64\degr2731 obtained with the SALT HRS on 2018 February 1. The 
principal lines and most prominent diffuse interstellar bands (DIBs) are indicated.}
\label{fig:spec}
\end{figure*}

The HRS is a dual beam, fibre-fed \'echelle spectrograph. In the MR mode it has a 2.23 arcsec diameter for both the object 
and sky fibres, providing a spectrum in the blue and red arms over the total spectral range of $\approx$3700--8900~\AA \,
with resolution of R$\sim$35\,000. For our observations both blue and red arms CCD were read out by a single amplifier with 
a 1$\times$1 binning. Three spectra of the ThAr arc-lamp and three spectral flats were obtained in this mode during a weekly 
set of HRS calibrations. Primary reduction of the spectra was performed with the SALT science pipeline (Crawford et al. 2010). 
The subsequent reduction steps were carried out using {\sc midas} HRS pipeline described in details in Kniazev, Gvaramadze \& 
Berdnikov (2016). A part of the spectrum obtained on 2018 February 1 is shown in Fig.\,\ref{fig:spec}. 

%%%%%%%%%%%%%%%%%%%%%%%%%%%%%%%%%%%%%%%%%%%%%%%%%%%%%%%%%%%%%%%%%%%%%%%%%%%
%%%%%%%%%%%%%%%%%%%%%%%%%%%%%%%%%%%%%%%%%%%%%%%%%%%%%%%%%%%%%%%%%%%%%%%%%%%

\section{Spectral analysis and stellar parameters}
\label{sec:ana}

\subsection{Classification of CPD$-$64\degr2731}
\label{sec:cla}

Figure\,\ref{fig:spec} shows that the spectrum of CPD$-$64\degr2731 is dominated by broad absorption lines of H, He\,{\sc i} 
and He\,{\sc ii}, indicating that we deal with a fast-rotating O star (cf. Vijapurkar \& Drilling 1993). Except of the weak
P\,Cygni profile of the He\,{\sc ii} $\lambda$4686 line, the only emission lines detected are those of N\,{\sc iii} 
$\lambda\lambda$4634, 4641--42, meaning that CPD$-$64\degr2731 is an O((f)) star (e.g. Sota et al. 2011). Using equivalent 
widths (EW) of the H\,{\sc i} $\lambda$4471 and He\,{\sc ii} $\lambda$4541 lines of $0.39\pm0.01$ \AA \, and $0.88\pm0.01$ 
\AA, respectively, one finds $\log W=\log$(EW(H\,{\sc i} 4471)/EW(He\,{\sc ii} 4541))=$-0.35\pm0.01$, which implies that 
CPD$-$64\degr2731 is an O5.5 star (Conti \& Alschuler 1971; see their table\,3). To determine the luminosity class, we use 
the criteria from Sota et al. (2014; see their table\,2), according to which CPD$-$64\degr2731 is a dwarf star (luminosity 
class V). To this classification we add a suffix `n', indicating that the projected rotational velocity of CPD$-$64\degr2731 
is $\approx300 \, \kms$ (see Section\,\ref{sec:mod}), so that CPD$-$64\degr2731 is an O5.5\,Vn((f)) star. 

\subsection{Spectral modelling}
\label{sec:mod}

\begin{figure*}
{\centering \resizebox*{2\columnwidth}{!}{\includegraphics[angle=0]{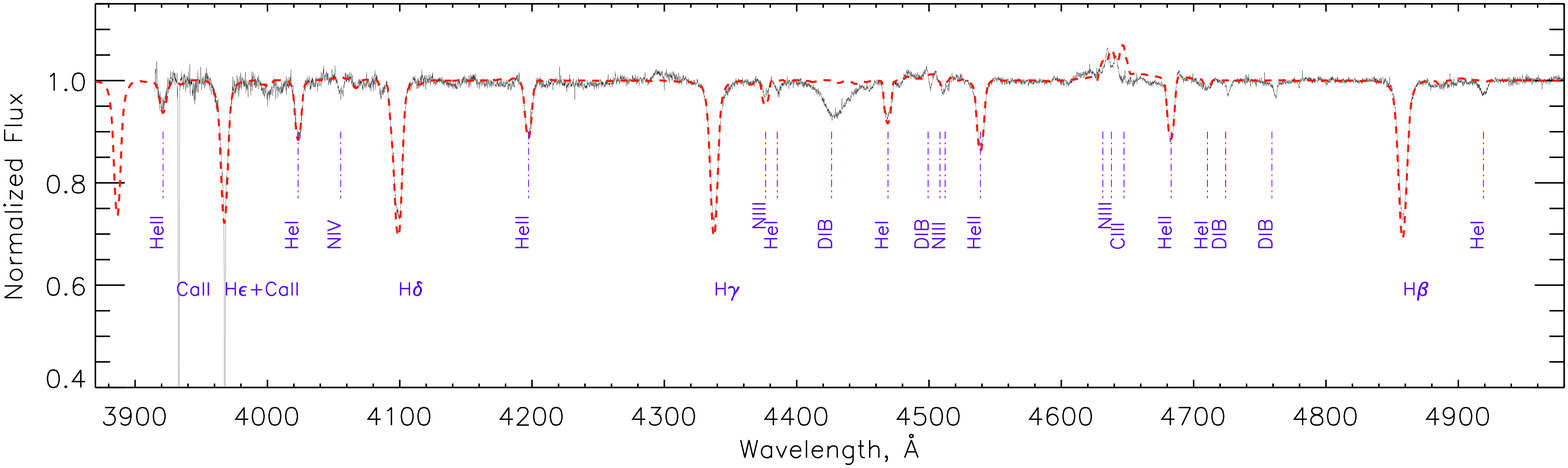}}}\\
{\centering \resizebox*{2\columnwidth}{!}{\includegraphics[angle=0]{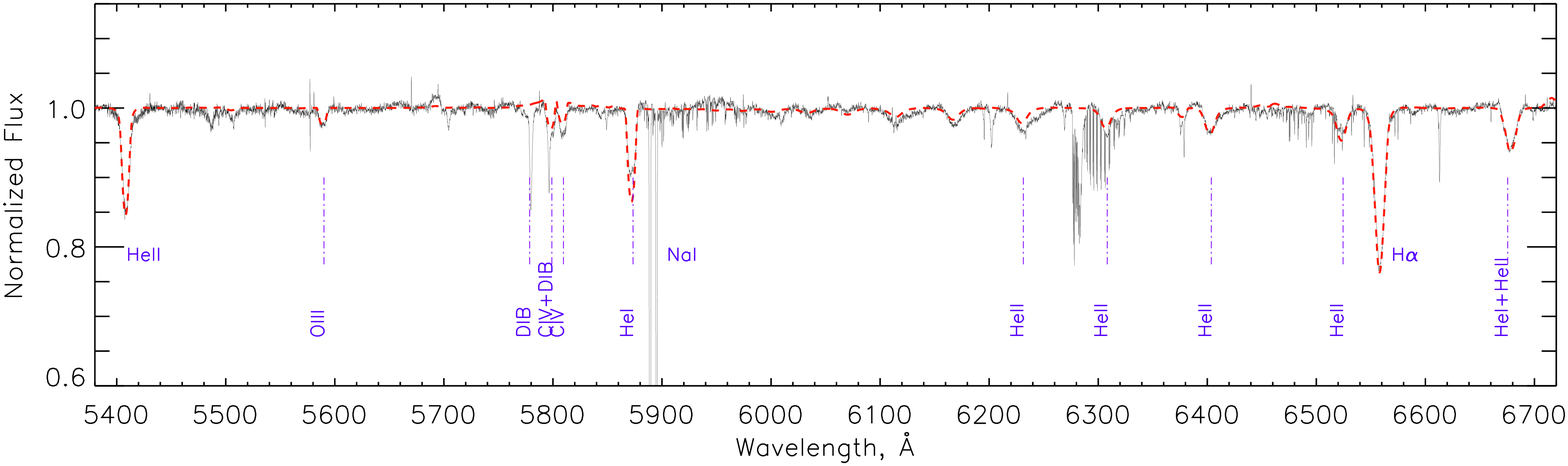}}}\\
\caption{Normalized HRS spectrum of CPD$-$64\degr2731 taken on 2018 February 1 (black solid line), compared with 
the best-fitting {\sc cmfgen} model (red dashed line) with the parameters as given in Table\,\ref{tab:par}.}
\label{fig:cmfgen}
\end{figure*}

To determine stellar parameters and chemical abundances of CPD$-$64\degr2731, we used the stellar atmosphere 
code {\sc cmfgen} (Hillier \& Miller 1998).
This code solves radiative transfer equations for objects with extended spherically symmetric outflows using 
either the Sobolev approximation or the full comoving-frame solution of the radiative transfer equation.  
{\sc cmfgen} incorporates line blanketing, the effects of Auger ionization and clumping. Every model is defined
by a hydrostatic stellar radius $R_*$, luminosity $L_*$, mass-loss rate $\dot{M}$, volume filling factor $f$, 
wind terminal velocity $v_\infty$, stellar mass $M_*$, and abundances $Z_i$ of chemical elements. 
Our modelling included the following elements: H, He, C, N, O, Si, S, P and Fe. The best-fitting
model is shown in Fig.\,\ref{fig:cmfgen}.

%%%%%%%%%%%%%%%%%%%%%%%%%%%%%%%%%%%%%%%%%%%%%%%%%%%%%%%%%%%%%%%%%%%%%%%%%%%
%%%%%%%%%%%%%%%%%%%%%%%%%%%%%%%%%%%%%%%%%%%%%%%%%%%%%%%%%%%%%%%%%%%%%%%%%%%

For the initial photospheric density structure we adopted the {\sc tlusty} hydrostatic model atmosphere of 
Hubeny \& Lanz (1995) and Lanz \& Hubeny (2003) for deep quasi-static layers connected to the wind with standard
$\beta$-velocity law just above the sonic point. We chose $\beta=0.8$ as the default value for our calculation, which 
is a typical value of this parameter for O dwarf stars (e.g. Repolust, Puls \& Herrero 2004). In the spectrum of 
CPD$-$64\degr2731 there are no lines with pronounced P\,Cyg profiles which can be used to measure $v_\infty$. We 
therefore assumed that $v_\infty=2.65v_{\rm esc}$ (Kudritzki \& Puls 2000), where $v_{\rm esc}$ is the escape velocity. 
The mass-loss rate was estimated by fitting profiles of the H$\alpha$ and He\,{\sc ii} $\lambda$4686 lines.

The clumping was described by the volume filling factor $f=\bar{\rho}/\rho(r)$, where $\bar{\rho}$ is the homogeneous 
(unclumped) wind density and ${\rho}$ is the density in clumps, which are assumed to be optically thin (Hillier \& Miller 
1999). The interclump medium is void. The filling factor depends on the radius as 
$f=f_\infty+(1-f_\infty)\exp(-v(r)/v_{\rm cl})$, where $f_\infty$ describes the density contrast and $v_{\rm cl}$ is a 
velocity at which the clumping becomes important. We adopted $v_{\rm cl}=10 \, \kms$ and $f_\infty=0.5$ to simultaneously 
fit the H$\alpha$ and He\,{\sc ii} $\lambda$4686 lines.

To derive the effective temperature $T_{\rm{eff}}$ (defined at the radius where the Rosseland optical depth 
equals 2/3), we compared intensities of the He\,{\sc i} and He\,{\sc ii} lines. The gravity was derived by fitting the 
wings of Balmer lines. 

The hydrogen and helium abundances were derived by iterative adjustment of the He-to-H abundance ratio and other 
fundamental parameters of CPD$-$64\degr2731 to reproduce the overall shape of all detected H and He lines.
The nitrogen abundance was estimated by analysing the behaviour of all nitrogen lines in the spectrum. The oxygen 
and carbon abundances were derived by use of the O\,{\sc iii} $\lambda$5592 and C\,{\sc iv} $\lambda\lambda$5801, 5812 
lines, respectively. Note that our best-fitting model shows the C\,{\sc iii} $\lambda\lambda4647-50-52$ lines in emission, 
while they are absent in the observed spectrum. Modelling of these lines is a difficult task because of very complex 
dependence of their profiles on the main stellar parameters ($\log g$, $T_{\rm eff}$, $\dot{M}$) as well as on the 
inclusion of other ions in calculations, such as Fe\,{\sc iv}, Fe\,{\sc v} and S\,{\sc iv} (Martins \& Hillier 2012).  
We suppose that the emergence of the C\,{\sc iii} $\lambda\lambda4647-50-52$ lines is connected with the atomic data 
we used in our model. For the abundances of Si, S, P and Fe we adopted solar values.

%%%%%%%%%%%%%%%%%%%%%%%%%%%%%%%%%%%%%%%%%%%%%%%%
\begin{table}
\caption{Comparison of the CNO abundances (by number) in CPD$-$64\degr2731 with the cosmic abundance standard (CAS; Nieva \&
Przybilla 2012) in the solar neighbourhood and initial abundances adopted in the evolutionary models by Brott et al. (2011).}
\label{tab:cno}
\begin{tabular}{lccc}
\hline
$\log$(X/H)+12 & CPD-64 2731             & CAS           & Brott et al. \\
\hline
C              & $8.30^{+0.18} _{-0.30}$ & $8.33\pm0.04$ & 8.13 \\
N              & $8.54^{+0.11} _{-0.15}$ & $7.79\pm0.04$ & 7.64 \\
O              & $8.30^{+0.18} _{-0.30}$ & $8.76\pm0.05$ & 8.55 \\
\hline                                                              
\end{tabular}
\end{table}
%%%%%%%%%%%%%%%%%%%%%%%%%%%%%%%%%%%%%%%%%%%%%%%%%%%%%%

In Table\,\ref{tab:cno}, we compare the CNO abundances (by number) of CPD$-$64\degr2731 with the cosmic abundance standard
(CAS) in the solar neighbourhood (Nieva \& Przybilla 2012) and the initial CNO abundances adopted in the evolutionary models 
of Galactic O stars by Brott et al. (2011). One can see that that the nitrogen abundance in CPD$-$64\degr2731 is 
enhanced by a factor of 6--8 (see also Section\,\ref{sec:blu}).

\subsection{Reddening, distance and luminosity of CPD$-$64\degr2731}
\label{sec:red}

\begin{figure*}
\includegraphics[width=14cm,angle=0]{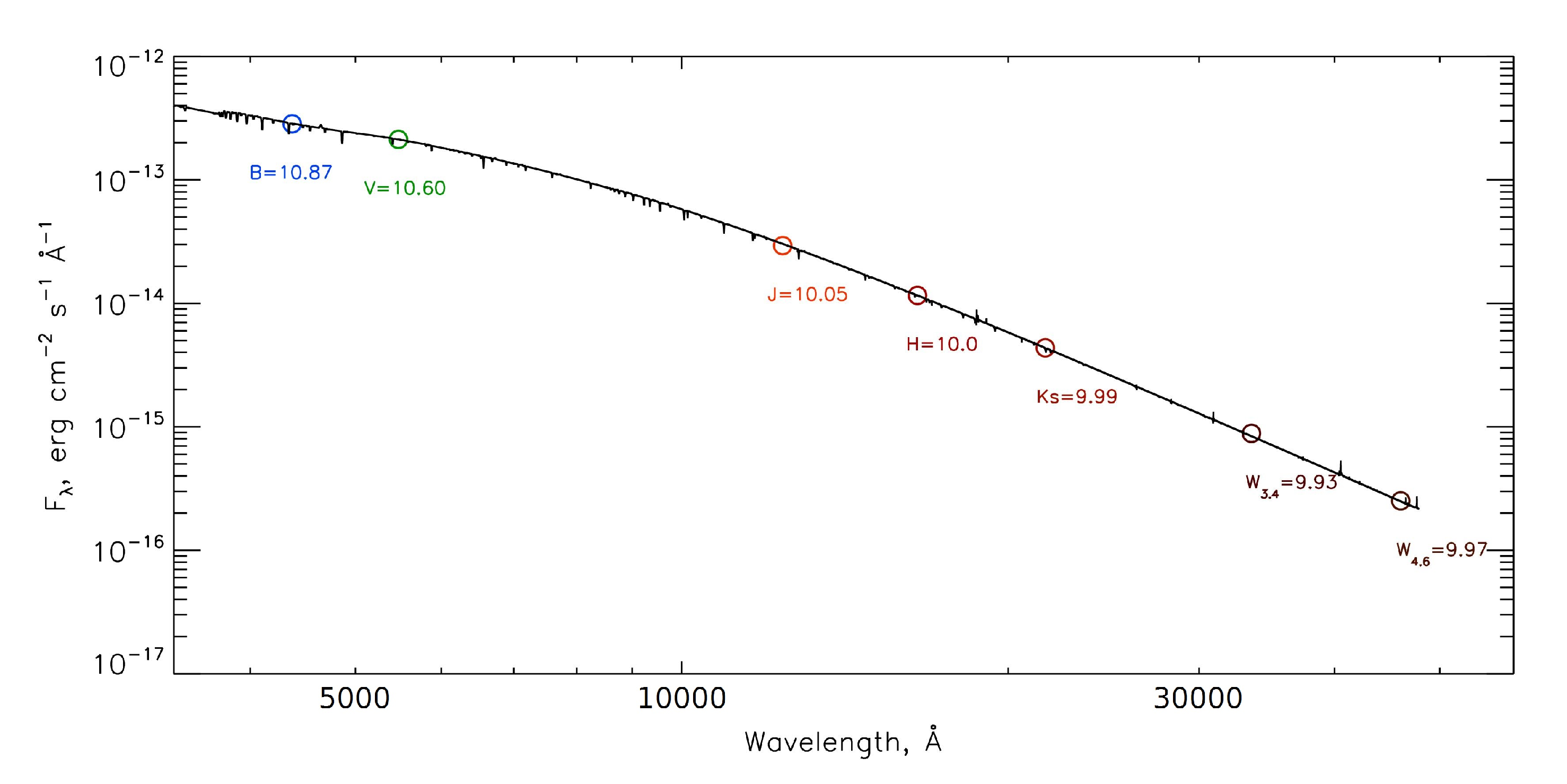}
\centering \caption{Observed flux distribution of CPD$-$64\degr2731 in absolute units, based on the photometric 
measurements (circles) compiled in Table\,\ref{tab:det}, compared to the emergent flux of the reddened model 
spectrum (black line) with the parameters as given in Table\,\ref{tab:par}.} 
\label{fig:sed}
\end{figure*}

To estimate the colour excess $E(B-V)$ towards CPD$-$64\degr2731, we compared the photometric measurements compiled in 
Table\,\ref{tab:det} with the spectral energy distribution (SED) in the model spectrum. We found 
that the slopes of the observed and model spectra match each other if $E(B-V)=0.56$ mag (Fig.\,\ref{fig:sed}). Note that 
the derived value of $E(B-V)$ is equal to the full Galactic reddening in the direction to CPD$-$64\degr2731 of 0.56 mag 
(Schlafly \& Finkbeiner 2011), which is expectable in view of the high Galactic latitude location of the star. Note also 
that the same reddening could be derived by using the $B$ and $V$ photometry of CPD$-$64\degr2731 and the intrinsic colour 
of the star of $(B-V)_0=-0.28$ mag (Martins \& Plez 2006), i.e. $E(B-V)=[(B-V)-(B-V)_0]=0.56$ mag.

%%%%%%%%%%%%%%%%%%%%%%%%%%%%%%%%%%%%%%%%%%%%%%%%%%%%%%%%%%%%%%%%%%%%%%%%%%%
%%%%%%%%%%%%%%%%%%%%%%%%%%%%%%%%%%%%%%%%%%%%%%%%%%%%%%%%%%%%%%%%%%%%%%%%%%%
%%%%%%%%%%%%%%%%%%%%%%%%%%%%%%%%%%%%%%%%%%%%%%%%
\begin{table*}
\caption{Stellar parameters of CPD$-$64\degr2731 for three distances (see text for details).}
\label{tab:par}
\begin{tabular}{lccc}
\hline
%\rule{0pt}{10pt}
$d$ (kpc)                      & 6.3            &  7.5           & 11 \\
$\log(L_*/\lsun)$              & $5.40\pm0.03$  & $5.57\pm0.03$  & $5.93\pm$0.01 \\
$R_* \, (\rsun)$               & 10             & 13             & 18 \\
$M_* \, (\msun)$               & 23             &  38            & 78 \\  
$v_\infty \, (\kms$)           & 2500           & 2900           & 3400 \\              
$\log(\dot{M}/\myr)$           & $-6.12\pm0.02$ & $-5.96\pm0.04$ & $-5.82\pm0.06$ \\ 
                               &                &                & \\
$T_{\rm eff}$ (kK)             & \MC {2}{c}{$40.0\pm1.0$}             \\
$\log g$                       & \MC {2}{c}{$3.8\pm0.1$}            \\   
$\beta$ (adopted)              & \MC {2}{c}{0.8}                   \\
$f$ (adopted)                  & \MC {2}{c}{0.5}                    \\  
$v\sin i \, (\kms)$            & \MC {2}{c}{$290^{+20} _{-40}$} \\
$v_{\rm h, hel} \, (\kms)$     & \MC {2}{c}{$-175.7\pm0.3$} \\
H (mass fraction)              & \MC {2}{c}{$0.71\pm0.04$}                     \\
He (mass fraction)             & \MC {2}{c}{$0.28\pm0.04$}                       \\
C (mass fraction)              & \MC {2}{c}{$(1.7\pm0.9)\times10^{-3}$}               \\
N  (mass fraction)             & \MC {2}{c}{$(3.5\pm1)\times10^{-3}$}              \\
O  (mass fraction)             & \MC {2}{c}{$(2.3\pm1.1)\times10^{-3}$}              \\
$E(B-V)$ (mag)                 & \MC {2}{c}{0.56} \\
\hline                                                              
\end{tabular}
\end{table*}
%%%%%%%%%%%%%%%%%%%%%%%%%%%%%%%%%%%%%%%%%%%%%%%%%%%%%%

With the known reddening to CPD$-$64\degr2731, one can estimate the distance to this star. Using the $V$-band absolute 
magnitude of an O5.5\,V star of $M_V=-5.14$ mag (Martins \& Plez 2006) and assuming the total-to-selective 
absorption ratio of $R=A_V/E(B-V)=3.1$, one finds the $V$-band extinction towards CPD$-$64\degr2731 of $A_V =1.74$ mag 
and the distance modulus of this star of DM$=V-M_V -A_V=14.00$ mag, implying a distance of $d\approx6.3$ kpc. 

The obtained distance estimate should be compared with the parallactic distance based on the data from the {\it Gaia}
second data release (DR2; Gaia Collaboration 2018). A simple inversion of the observed parallax of CPD$-$64\degr2731 of 
$0.091\pm0.026$ mas places this star at a distance of $d=10.95^{+4.36} _{-2.42}$ kpc. A shorter distance 
to CPD$-$64\degr2731 of $d=7.54^{+1.67} _{-1.21}$ was derived from the DR2 data by Bailer-Jones et al. (2018) by using a 
Bayesian prior knowledge approach. This distance estimate agrees within the error margins with that based on the 
$M_V$-spectral type calibration. 

To determine the luminosity, we recomputed the fluxes for three values of $d=6.3,7.5$ and 11 kpc. The resulting 
fluxes were corrected for the interstellar extinction, and then compared to the model spectra convolved with the transmission 
curves of the standard $B$ and $V$ filters. The obtained luminosity and other stellar parameters derived for three
above values of $d$ are listed in Table\,\ref{tab:par} along with the surface abundances of main chemical elements (by mass).
Note that at $d=6.3$ kpc (11 kpc) the inferred mass of CPD$-$64\degr2731 of $23 \, \msun$ ($78 \, \msun$) is much smaller
(larger) than what one might expect for an unevolved O5.5 star (cf.  Martins, Schaerer \& Hillier 2005; Weidner \& Vink 2010). 
Proceeding from this, in what follows we adopt the distance to CPD$-$64\degr2731 of 7.5 kpc, and note that the main 
conclusions of the paper do not change if the actual distance to the star is within reasonable limits around this value.

\subsection{Rotational and radial velocities}
\label{sec:rad}

The high resolution of the HRS spectra allowed us to determine the projected rotational velocity of CPD$-$64\degr2731
and to search for possible radial velocity variability of this star.

The projected rotational velocity was derived through the Fourier transform of the He\,{\sc i} $\lambda$4471 line profile 
using the {\sc iacob-broad} code (Sim\'on-D\'iaz \& Herrero 2007, 2014). The obtained value of $v\sin i\approx290^{+20} _{-40}$, 
where $i$ is the angle between the stellar rotation axis and our line-of-sight, turns out to be extremely high for a O5.5 star
(we further discuss this point in Section\,\ref{sec:blu}).

The heliocentric radial velocity, $v_{\rm r,hel}$, of CPD$-$64\degr2731 given in Table\,\ref{tab:par} is the mean value 
based on three HRS spectra. The measurements were done with a dedicated software package developed by our team (Katkov 
et al., in preparation). This software is based on the library of high-resolution theoretical stellar spectra (Coelho 2014) 
and is designed to derive radial velocities for components of binary systems. We simultaneously fitted the available observed 
spectra using the same model spectrum interpolated from the grid of stellar models and convolved with the instrumental 
resolution and $v\sin i$, and shifted with individual line-of-sight velocity for a given epoch. In clear cases of binary stars 
the package uses two model spectra with individual velocities. We did not reveal any signature of a second star in the spectra 
of CPD$-$64\degr2731, and therefore measured only one radial velocity for each analysed spectrum. The obtained values of 
$v_{\rm r,hel}$ of $-178.6\pm0.4 \, \kms$ (2018 February\,1), $-174.9\pm0.6 \, \kms$ (2018 February\,25) and 
$-173.7\pm0.6 \, \kms$ (2018 July 18) indicate that CPD$-$64\degr2731 might be a wide binary system (cf. Section\,\ref{sec:blu}), 
unless the uncertainties in the radial velocity measurements are underestimated due to possible unaccounted systematic errors.  

%%%%%%%%%%%%%%%%%%%%%%%%%%%%%%%%%%%%%%%%%%%%%%%%%%%%%%%%%%%%%%%%%%%%%%%%%%%
\begin{figure}
\includegraphics[width=8.5cm,angle=0]{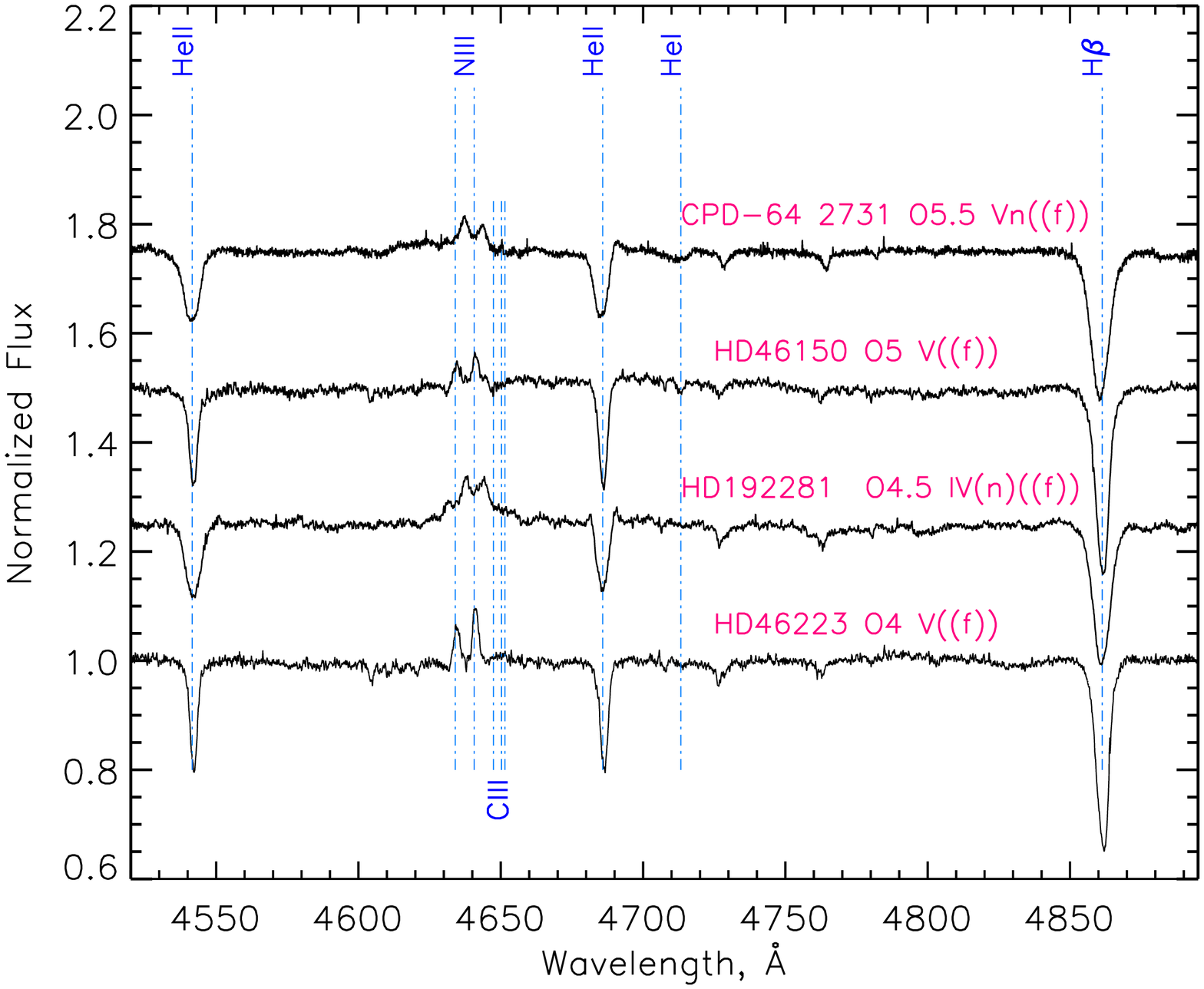}
\centering \caption{Comparison of a portion of the HRS spectrum of CPD$-$64\degr2731 with the corresponding parts of spectra 
of three other O((f)) stars. Note that the N\,{\sc iii} lines in the spectra of CPD$-$64\degr2731 and HD\,192281 are shifted 
redwards.} 
\label{fig:shi}
\end{figure}
%%%%%%%%%%%%%%%%%%%%%%%%%%%%%%%%%%%%%%%%%%%%%%%%%%%%%%%%%%%%%%%%%%%%%%%%%%%

Finally, we note that the only emission lines detected in the spectrum of CPD$-$64\degr2731, namely the N\,{\sc iii} 
$\lambda\lambda$4634, 41--42 ones, are shifted redwards with respect to the absorption lines by $\approx4$\,\AA \, or $\approx260 
\, \kms$. Similar shift of the same lines was also found in the spectrum of the O4.5\,IV(n)((f)) (Ma\'iz Apell\'aniz et al. 2016) 
star HD\,192281 (Sota et al. 2014), as illustrated in Fig.\,\ref{fig:shi}, where we compare a portion of the spectrum of 
CPD$-$64\degr2731 with the corresponding parts of spectra of HD\,192281 and two other O((f)) stars (retrieved from the ELODIE 
archive at Observatoire de Haute-Provence; Moultaka et al. 2004). Sota et al. (2014) suggested that the shift of the N\,{\sc iii}
lines in the spectrum of HD\,192281 is caused by the binarity of this star, which is also responsible for the line broadening in 
the spectrum. The possible binary status of HD\,192281 was first suggested by Barannikov (1993) who found small-amplitude radial 
velocity variability with a period of 5.48 day. More recent measurements by Cazorla et al. (2017) also found evidence for 
variability, but did not confirm its periodicity, meaning that HD\,192281 might not necessarily be a binary system.
Since the N\,{\sc iii} lines are produced in the stellar photosphere (Brucato \& Mihalas 1971), we speculate that their shift is 
somehow related to asymmetry of the photosphere, probably caused by the fast rotation and/or strong magnetic field (expected in 
the star if it is a binary merger product; cf. Section\,\ref{sec:blu}). Spectropolarimetric observations of CPD$-$64\degr2731 
could be of high importance to clarify this issue.

%%%%%%%%%%%%%%%%%%%%%%%%%%%%%%%%%%%%%%%%%%%%%%%%%%%%%%%%%%%%%%%%%%%%%%%%%%%
\begin{table*}
\caption{Astrometric and kinematic data on CPD$-$64\degr2731 (see text for details).}
\label{tab:pro}
\begin{tabular}{cccccccc}
\hline 
$d$   & $v_{\rm l}$   & $v_{\rm b}$    & $v_{\rm r}$    & $v_{\rm tr}$  & $v_*$         & $z$  & $t_{\rm kin}$ \\
(kpc) & ($\kms$)      & ($\kms$)       & ($\kms$)       & ($\kms$)      & ($\kms$)      & (pc) & (Myr) \\
\hline 
6.3   & $-18.2\pm1.0$ & $-70.4\pm1.1$  & $-125.2\pm0.3$ & $72.7\pm1.1$  & $144.8\pm0.6$ & 410  & 5.9 \\
7.5   & $-20.4\pm1.2$ & $-86.6\pm1.3$  & $-136.0\pm0.3$ & $89.0\pm1.3$  & $162.5\pm0.8$ & 490  & 5.7 \\
11    & $-68.3\pm1.8$ & $-134.2\pm1.9$ & $-186.2\pm0.3$ & $150.6\pm1.5$ & $239.5\pm1.0$ & 720  & 5.4 \\
\hline
\end{tabular}
\end{table*}
%%%%%%%%%%%%%%%%%%%%%%%%%%%%%%%%%%%%%%%%%%%%%%%%%%%%%%%%%%%%%%%%%%%%%%%%%%%

%%%%%%%%%%%%%%%%%%%%%%%%%%%%%%%%%%%%%%%%%%%%%%%%%%%%%%%%%%%%%%%%%%%%%%%%%%%
\begin{figure*}
\includegraphics[width=14cm,angle=0]{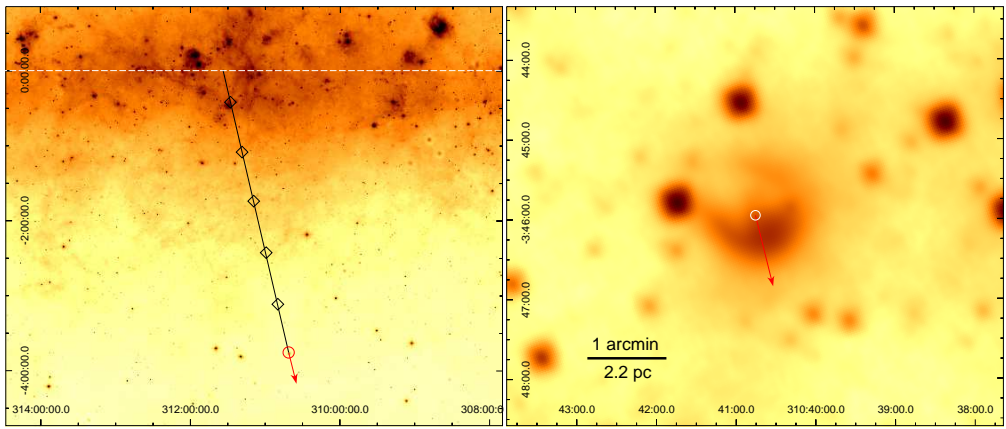}
\centering \caption{Left-hand panel: {\it WISE} 22\,\micron \, image of the field containing CPD$-$64\degr2731 and its 
surrounding nebula (marked by a circle). The arrow shows the direction of peculiar transverse velocity of CPD$-$64\degr2731, 
derived using the {\it Gaia} DR2 proper motion measurement and adopting a distance to this star of 7.5 kpc.
The solid (black) line shows the trajectory of CPD$-$64\degr2731, and diamonds mark the positions of the star 1, 2, 3, 4 and 
5 Myr ago. The Galactic plane is indicated by a white dashed line. Right-hand panel: {\it WISE} 12\,\micron \, image of the 
nebula with position of CPD$-$64\degr2731 marked by a circle. The arrow shows the direction of peculiar transverse velocity 
of the star. In both panels, the coordinates (in units of degrees) are the Galactic longitude and latitude on the horizontal 
and vertical scales, respectively. At a distance of 7.5 kpc, 1\degr and 1 arcmin correspond, respectively, to $\approx130$ 
and 2.2 pc. See the text for details.} 
\label{fig:gal}
\end{figure*}
%%%%%%%%%%%%%%%%%%%%%%%%%%%%%%%%%%%%%%%%%%%%%%%%%%%%%%%%%%%%%%%%%%%%%%%%%%%

\section{CPD$-$64\degr2731 as a runaway star}
\label{sec:run}

The {\it Gaia} DR2 provides high-precise proper motion measurements for CPD$-$64\degr2731 (see Table\,\ref{tab:pro}):
$\mu _\alpha \cos \delta=-6.69\pm0.03 \, {\rm mas} \, {\rm yr}^{-1}$ and $\mu _\delta=-4.83\pm0.04 \, {\rm mas} \, 
{\rm yr}^{-1}$. Using these data and $v_{\rm r,hel}$ from Table\,\ref{tab:par}, we calculated the peculiar transverse velocity 
$v_{\rm tr}=(v_{\rm l}^2 +v_{\rm b}^2)^{1/2}$, where $v_{\rm l}$ and $v_{\rm b}$ are, respectively, the peculiar velocity components 
along the Galactic longitude and latitude, the peculiar radial velocity $v_{\rm r}$, and the total space velocity $v_*$ of the 
star. For this, we adopted the solar Galactocentric distance of $R_0 = 8.0$ kpc and the circular Galactic rotation velocity of 
$\Theta _0 =240 \, \kms$ (Reid et al. 2009), and the solar peculiar motion of $(U_{\odot},V_{\odot},W_{\odot})=(11.1,12.2,7.3) 
\, \kms$ (Sch\"onrich, Binney \& Dehnen 2010). For the error calculation, only the errors of the proper motion and the radial 
velocity measurements were considered. For the sake of completeness, we calculated velocities for three value of distance:
$d=6.3$, 7.5 and 11 kpc. The results are given in Table\,\ref{tab:pro} along with the separation of CPD$-$64\degr2731 from the 
Galactic plane, $z=d\sin b$, and the kinematic age of this star, $t_{\rm kin}=z/v_{\rm b}$.

From Table\,\ref{tab:pro} it follows that CPD$-$64\degr2731 is a high-velocity  runaway star. The runaway status of 
CPD$-$64\degr2731 is also supported by the large separation of this star from the Galactic plane (cf. Blaauw 1961; van Oijen 1989),
which is much larger than the scale heights of runaway O and `normal' OB stars of, respectively, $\approx100$ pc (Stone 1979) and 
$\approx50$ pc (Stone 1979; Reed 2000).

At the adopted distance of $d=7.5$ kpc, the space velocity of CPD$-$64\degr2731 is $\approx160 \, \kms$, the star is
approaching us with a velocity of $\approx140 \, \kms$, and the vector of its total space velocity is inclined to our line 
of sight by $\approx33$\degr. Although a $40 \, \msun$ star can, in principle, attain a velocity of $160 \, \kms$ in the course 
of dissolution of a close binary system following asymmetric supernova explosion (e.g. Portegies Zwart 2000; Eldridge, Langer 
\& Tout 2011), the young age of CPD$-$64\degr2731 ($\la3$\,Myr; see next section) and the high rotational velocity of this star 
make this possibility unlikely (see below). This, along with observations that massive stars form in a clustered way (Lada \& 
Lada 2003; Gvaramadze et al. 2012b), indicates that CPD$-$64\degr2731 attained its high space velocity because of a three- or 
four-body dynamical encounter with other massive stars in the parent star cluster (Poveda, Ruiz \& Allen 1967; Leonard \& Duncan 
1990). Furthermore, the high mass and velocity of CPD$-$64\degr2731 imply that the kinetic energy of this star is one of 
the largest among known runaway stars\footnote{Two other massive runaway stars with record kinetic energy are the
O2 stars Sk$-$67\degr22 and BI\,237 in the Large Magellanic Cloud (Massey et al. 2005).}, and that the stars involved in 
the dynamical interaction were much more massive than this star.

For example, numerical experiments show (see figs 7 and 11 in Gvaramadze \& Gualandris 2011) that a $40 \, \msun$ star could 
be accelerated to a velocity of $\geq160 \, \kms$ in the course of encounter with a massive binary of total mass similar to 
that of the most massive known binaries in the Milky Way, WR\,20a and NGC\,3603-A1, whose masses are, respectively, 
$\approx160$ and $200 \, \msun$. Both WR\,20a and NGC\,3603-A1 are members of young, very massive ($\sim10^4 \, \msun$) star 
clusters, which, like other star clusters of this mass, are located in the Galactic plane. It is likely, therefore, that 
CPD$-$64\degr2731 was ejected from a similar very massive cluster in the Galactic plane.

Figure\,\ref{fig:gal} shows that CPD$-$64\degr2731 is moving almost perpendicular to the Galactic plane and that its past 
trajectory intersects the Galactic plane at $l\approx311\fdg5$. Note that the orientation of the transverse 
peculiar velocity of the star only slightly depends on the distance; by varying the distance between 6 and 9 kpc, we found 
that the corresponding past trajectories of CPD$-$64\degr2731 intersect the Galactic plane within a half degree from each other.
In this direction our line of sight first crosses the Carina-Sagittarius arm at $\approx1.5$ kpc and then
tangentially enters into the Crux-Scutum spiral arm at $\approx5$ kpc (Cordes \& Lazio 2002). All distance estimates for 
CPD$-$64\degr2731 imply that this star was formed in the Crux-Scutum arm.

We searched for known young ($\leq10$ Myr) star clusters in the region between Galactic longitudes 310\degr and 313\degr 
using the WEBDA data base\footnote{http://webda.physics.muni.cz/} (Mermilliod 1995), but did not find any. This non-detection 
may indicate that the parent cluster to CPD$-$64\degr2731 is already dissolved following residual-gas expulsion at the very 
beginning of cluster dynamical evolution (Tutukov 1978) or is heavily obscured by the foreground dust in the Galactic plane.
Note that the very massive binary involved in the dynamical interaction with CPD$-$64\degr2731 and recoiled in the opposite
direction to this star should have ended its life in supernova explosions several Myr ago.

%%%%%%%%%%%%%%%%%%%%%%%%%%%%%%%%%%%%%%%%%%%%%%%%%%%%%%%%%%%%%%%%%%%%%%%%%%%
\begin{figure}
\includegraphics[width=8.5cm,angle=0]{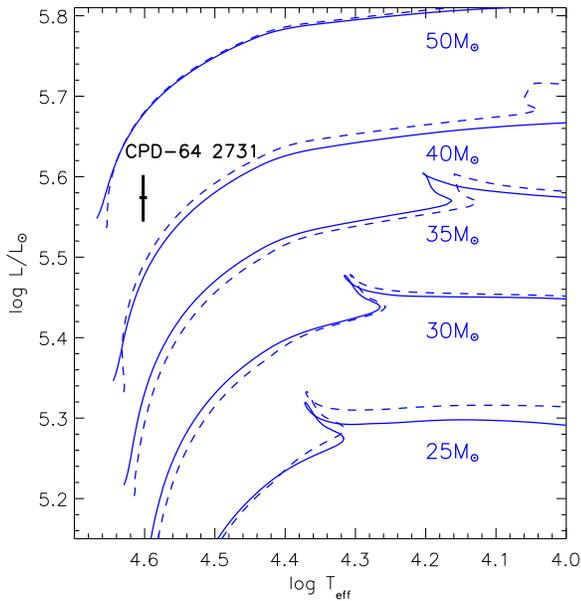}
\centering \caption{Position of CPD$-$64\degr2731 in the Hertzsprung-Russell diagram (cross with error bars). Also shown 
are the evolutionary tracks of Brott et al. (2011) for Galactic O stars. The dashed and solid lines correspond, respectively, 
to the tracks with initial rotational velocities of 0 and $\approx400 \, \kms$.}
\label{fig:hr}
\end{figure}
%%%%%%%%%%%%%%%%%%%%%%%%%%%%%%%%%%%%%%%%%%%%%%%%%%%%%%%%%%%%%%%%%%%%%%%%%%%

%%%%%%%%%%%%%%%%%%%%%%%%%%%%%%%%%%%%%%%%%%%%%%%%%%%%%%%%%%%%%%%%%%%%%%%%%%%
\begin{table*}
\caption{Comparison of CPD$-$64\degr2731 with model stars from Brott et al. (2011).}
\label{tab:bro}
\renewcommand{\footnoterule}{}
\begin{minipage}{\textwidth}
\begin{center}
\begin{tabular}{lcccccc}
\hline 
                              & CPD$-$64\degr2731 & Model 1 & Model 2 & Model 3 & Model 4  & Model 5 \\
\hline
$M_{\rm init} (\msun)$        & --                & 40      & 40      & 40      & 40       & 50 \\
$v_{\rm init} (\kms)$         & --                & 0       & 317     & 367     & 464      & 364    \\     
age (Myr)                     & --                & 2.1     & 2.15    & 2.8     & 4.0      & 2.4     \\
$M_* (\msun)$                 & 38                & 38.4    & 38.2    & 37.4    & 34.9     & 45.4    \\
$T_{\rm eff}$ (kK)            & 39.5--40.5        & 40      & 40      & 40      & 40       & 40  \\
$\log(L_*/\lsun)$             & 5.54--5.60        & 5.47    & 5.45    & 5.49    & 5.58     & 5.68   \\
$R_* (\rsun)$                 & 13                & 11      & 11      & 12      & 13       & 14     \\
$\log g$                      & 3.7--3.9          & 3.9     & 3.9     & 3.9     & 3.8      & 3.8 \\
$v_{\rm surf} (\kms)$         & 250--310$^a$      & 0       & 283     & 300     & 282      & 275   \\  
He (mass fraction)            & 0.24--0.32        & 0.26    & 0.31    & 0.30    & 0.27     & 0.28   \\
C (mass fraction) (10$^{-3}$) & 0.80--2.60        & 1.18    & 0.63    & 0.23    & 0.08     & 0.32 \\
N (mass fraction) (10$^{-3}$) & 2.50--4.50        & 0.44    & 1.84    & 3.68    & 5.02     & 3.14   \\ 
O (mass fraction) (10$^{-3}$) & 1.20--3.40        & 4.13    & 3.29    & 1.71    & 0.37     & 2.20    \\
\hline
\end{tabular}
\end{center}
\end{minipage}
{\it Note.} $^a v\sin i$.
\end{table*}
%%%%%%%%%%%%%%%%%%%%%%%%%%%%%%%%%%%%%%%%%%%%%%%%%%%%%%%%%%%%%%%%%%%%%%%%%%% 

\section{CPD$-$64\degr2731 as a spun-up and rejuvenated star}
\label{sec:blu}

Figure\,\ref{fig:hr} shows the position of CPD$-$64\degr2731 in the Hertzsprung-Russell diagram along with evolutionary 
tracks for Galactic non-rotating and rotating O stars by Brott et al. (2011). From the figure it follows that if 
CPD$-$64\degr2731 was born as a single star, then its initial mass, $M_{\rm init}$, was between 40 and $50 \, \msun$. 

In Table\,\ref{tab:bro}, we compile the main parameters of CPD$-$64\degr2731 and the corresponding parameters of several 40 
and $50 \, \msun$ model stars from Brott et al. (2011) for the times when their effective temperatures decrease to 40\,kK. 
One can see that in the non-rotating model star (model\,1) the CNO abundances remain almost equal to the initial abundances 
adopted in Brott et al. (2011). With the increase of the initial rotational velocity, $v_{\rm init}$, the nitrogen abundance 
in the model stars also grows and becomes comparable to that in CPD$-$64\degr2731 for $v_{\rm init}\approx400 \, \kms$ 
(models\,3 and 5). For this $v_{\rm init}$ the oxygen abundance in the model stars also agrees with that derived for 
CPD$-$64\degr2731, while the carbon abundance is much lower, which in part could be due to low initial abundance of this 
element adopted in the models (see Table\,\ref{tab:cno}), and in part due to inaccuracy of our spectral modelling. For higher 
values of $v_{\rm init}$ (model\,4), the model star remains hotter than 40 kK for a longer time, and, correspondingly, its 
N abundance increases even more, while the C and O abundances become much more depleted as compared to those in 
CPD$-$64\degr2731. 

More importantly, Table\,\ref{tab:bro} shows that if CPD$-$64\degr2731 underwent effectively single-star evolution then its age 
should be $\la3$\,Myr, i.e. a factor of two younger than the kinematic age (see Table\,\ref{tab:pro}). This implies that 
CPD$-$64\degr2731 could be a rejuvenated and spun-up product of binary mass transfer or the result of a binary merger, i.e., a 
massive blue straggler. This in turn implies that CPD$-$64\degr2731 was initially a compact binary system, which was ejected via 
strong few-body dynamical interaction (e.g. Oh \& Kroupa 2012). Numerical scattering experiments by Kroupa (1998) showed that a 
binary system can survive the ejection process only if its space velocity is less than the orbital velocity. These experiments 
also showed that the higher ejection velocity of runaway binaries the closer their mass ratio to unity.

Let us assume that the progenitor binary of CPD$-$64\degr2731 was composed of two $\approx20 \, \msun$ stars. Such a binary
would undergo stable mass transfer if its initial orbital period was $\ga3$\,d (Wellstein, Langer \& Braun 2001) or merger if 
the orbital period was shorter and the masses of the binary components were equal to each other (cf. Schneider et al. 2015).
The enhanced nitrogen abundance in CPD$-$64\degr2731 could be the direct result of binary mass transfer (Langer 2012) or of
merger process of two stars (Glebbeek et al. 2013). The merger ends with a single star, while in the case of mass transfer one
would expect a helium star companion in a wide ($\approx30$\,d) orbit around CPD$-$64\degr2731, which can be quite faint at 
optical wavelengths to be detected (e.g. G\"otberg, de Mink \& Groh 2017). In the latter case, CPD$-$64\degr2731 should show 
a radial velocity variation of $\approx20 \, \kms$ (Wellstein et al. 2001). Since the existing data do not allow us to firmly
assert that CPD$-$64\degr2731 is a binary system, we reserve the possibility that this star could be a binary merger product. 

The high projected rotational velocity of CPD$-$64\degr2731 supports the possibility that this star is a product of 
binary evolution. Indeed, observations of $\approx200$ Galactic OB stars show that the distribution of their projected 
rotational velocities peaks at $\approx50 \, \kms$ and that only few of them have velocities of $\ga 300 \, \kms$ 
(Sim\'on-D\'iaz \& Herrero 2014). Furthermore, only two of these fast-rotating stars are of spectral type earlier than O6.
Similarly, the study of projected rotational velocities of $\approx200$ apparently single O stars in the 30 Doradus star 
forming region of the Large Magellanic Cloud by Ram\'irez-Agudelo et al. (2013) shows that none of the early O-type (O2--5) 
stars rotate faster than $300 \,  \kms$. The lack of fast rotators among the early O-type stars could be explained by removal 
of angular momentum of their surface layers through wind mass loss, which is more efficient for massive stars because of their 
strong winds (e.g. Langer 1998). On the other hand, the existence of the high-velocity tail in the distribution of rotational 
velocities could be understood if it is populated by products of binary interaction, e.g. by spun-up mass gainers or 
merged binaries (Sana et al. 2012; de Mink et al. 2013).

Interestingly, observations also show that the fraction of fast-rotating O stars is larger outside star clusters
(Ram\'irez-Agudelo et al. 2013; see their fig.\,12). Ram\'irez-Agudelo et al. (2013) suggested that the fast rotators in 
the field may be spun-up post-binary interaction products that were ejected from parent clusters when their primary stars 
exploded in supernovae. Alternatively, the rapid rotation of at least some of the field O stars could be explained 
by mass transfer in or merger of runaway binaries (recall that most runaways are binary systems, meaning that they were 
ejected in the field because of dynamical few-body interactions; Chini et al. 2012). This possibility is supported by the 
results of numerical modelling of young massive star clusters, which show that the orbital period distribution of runaway 
binaries is biased towards shorter periods as compared to binaries retained in the clusters (Oh \& Kroupa 2016), and that 
the runaway binaries tend to have eccentric orbits (Leonard \& Duncan 1990), implying that they are more prone for binary 
interaction processes. 

Finally, we point out that within the merger scenario, the high rotational velocity of CPD$-$64\degr2731 serves as an 
indication that the merger having occurred very recently. Indeed, if the origin of magnetic field in massive stars is due to binary 
merger (e.g. Ferrario et al. 2009; Langer 2012; see also Section\,\ref{sec:con}) then the spin-down of the merger product is 
extremely fast, such that most magnetic stars rotate very slowly (Fossati et al. 2016).

\section{The origin of the nebula}
\label{sec:ori}

The high space velocity of CPD$-$64\degr2731 implies that the ram pressure of the ISM should play important role in
shaping the nebula. This inference is supported by the orientation of the peculiar transverse velocity vector, which 
is pointed towards the brightest side of the nebula, and by the offset of the star from the centre of the nebula in the 
same direction. Below we propose a possible scenario for the origin of the nebula.

First, we note that the nebula hardly can be explained in the framework of single star evolution. Indeed, the unevolved status
of CPD$-$64\degr2731 excludes the possibility that the nebula was formed because of the wind-wind interaction at the advanced 
stages of stellar evolution, as it is the case for circumstellar nebulae around evolved massive stars (e.g. Garc\'ia-Segura 
et al. 1996a,b; Brighenti \& D'Ercole 1997). Also, the high temperature of CPD$-$64\degr2731 implies that the nebula cannot be 
formed during an episode of enhanced mass loss caused by the bi-stability jump -- a factor of 10 increase of the mass-loss rate 
when the stellar temperature drops below a critical value of $\approx21$\,kK (Pauldrach \& Puls 1990; Vink 2018). Furthermore, 
the general appearance of the nebula and the nearly central location of CPD$-$64\degr2731 within it allow us to argue that the 
nebula is not a genuine bow shock (we validate this assertion in Section\,\ref{sec:num} by means of numerical modelling).

We concluded above that CPD$-$64\degr2731 is either a mass gainer from a binary mass transfer or a merger star.
In both cases, a few solar masses of material are likely spilled and blown into the circumstellar environment with rather small 
velocity (e.g., Petrovic et al. 2005; Glebbeek et al. 2013). Therefore, the nebula could be caused by changes in the mass-loss 
rate and wind velocity following the binary interaction. We suggest the following scenario. 

Originally, CPD$-$64\degr2731 was a (runaway) binary system composed of two almost equal-mass stars. The combined 
winds from these stars created a bow shock ahead of the system. The stand-off distance of the bow shock, i.e. the distance 
between the stars and the nearest side of the bow shock (apex), is given by (Baranov, Krasnobaev \& Kulikovskii 1970)
\begin{equation}
R_0 =\left({\dot{M}v_\infty\over 4\pi\rho_{\rm ISM}v_* ^2}\right)^{1/2} \, ,
\label{eqn:ram}
\end{equation}
where $\rho_{\rm ISM}=1.4m_{\rm H}n_{\rm ISM}$, $m_{\rm H}$ is the mass of a hydrogen atom, and $n_{\rm ISM}$ is
the number density of the local ISM. At some point, the binary interaction starts. This leads to a strong and slow wind,
either from the fast rotating accretion star or from the bloated merger product. In both cases, this wind is expected to last 
for a Kelvin-Helmholtz time-scale (about several thousand years), after which the binary product will establish a fast wind 
started to blow a bubble in the ambient medium. The geometry of the bubble is affected by the presence of shocked ISM ahead 
of the star and the bubble initially looks like a horseshoe-shaped nebula with its open ends pointing downstream and the star 
located close to the geometric centre of the nebula. We suggest that currently we are witnessing just this situation and that 
later on, after the star will completely overrun the preprocessed ISM, the bubble will transform into a classical bow shock. 
To validate this scenario, we performed numerical simulations, as described and discussed in the next section.

Our scenario for the formation of the observed circumstellar nebula around CPD$-$64\degr2731 has the implication that the 
strong binary interaction happened only recently. This in turn strongly supports the idea that CPD$-$64\degr2731 
was dynamically ejected from a dense star cluster as a binary system, and argues against the possibility that it was ejected 
from a broken-up binary as a consequence of the explosion of its companion.

\section{Numerical simulations}
\label{sec:num}

%%%%%%%%%%%%%%%%%%%%%%%%%%%%%%%%%%%%%%%%%%%%%%%%%%%%%%%%%%%%%%%%%%%%%%%%%%%
\begin{figure*}
\includegraphics[width=12cm,angle=0]{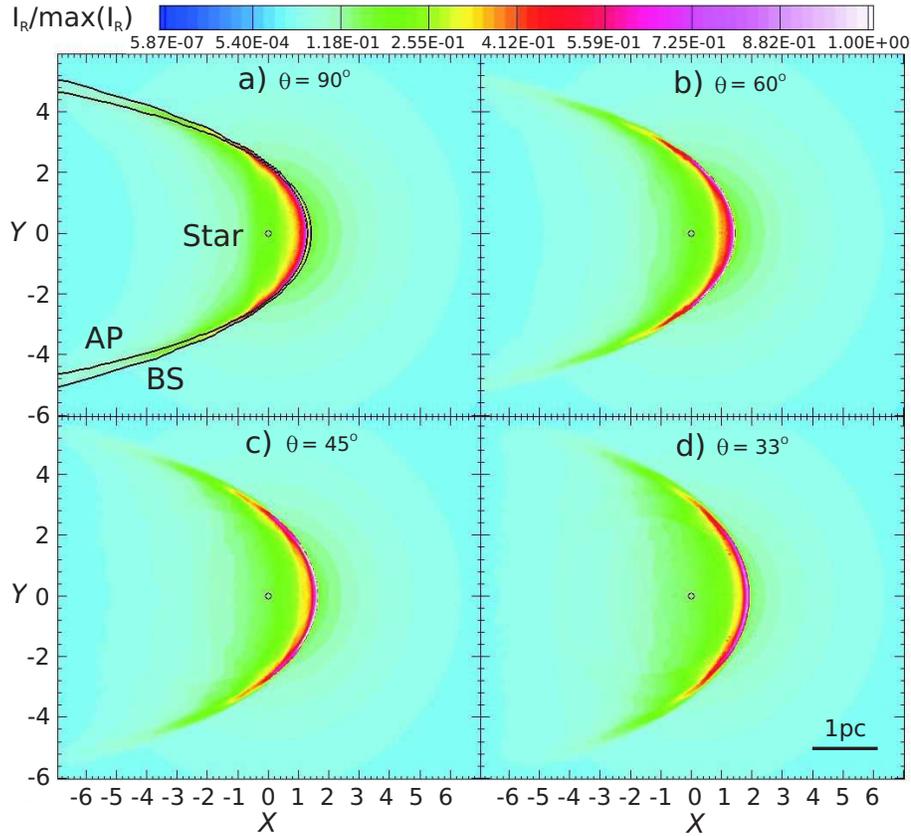}
\centering \caption{Synthetic maps of 22\,\micron \, dust emission (normalized to the maximum surface brightness of panel d) 
from the model bow shock at four inclination angles of the line of sight to the symmetry axis of the simulation ($Y=0$): 
a) 90\degr, b) 60\degr, c) 45\degr, and d) 33\degr. The wind parameters of the wind-blowing star are equal to those of 
CPD$-$64\degr2731. Solid lines in panel\,a) show the positions of astropause (AP) and bow shock (BS). The distance units on 
the $X$ and $Y$ axes are in 0.45 pc.} 
\label{fig:bow}
\end{figure*}
%%%%%%%%%%%%%%%%%%%%%%%%%%%%%%%%%%%%%%%%%%%%%%%%%%%%%%%%%%%%%%%%%%%%%%%%%%%

%%%%%%%%%%%%%%%%%%%%%%%%%%%%%%%%%%%%%%%%%%%%%%%%%%%%%%%%%%%%%%%%%%%%%%%%%%%
\begin{figure*}
\includegraphics[width=17cm,angle=0]{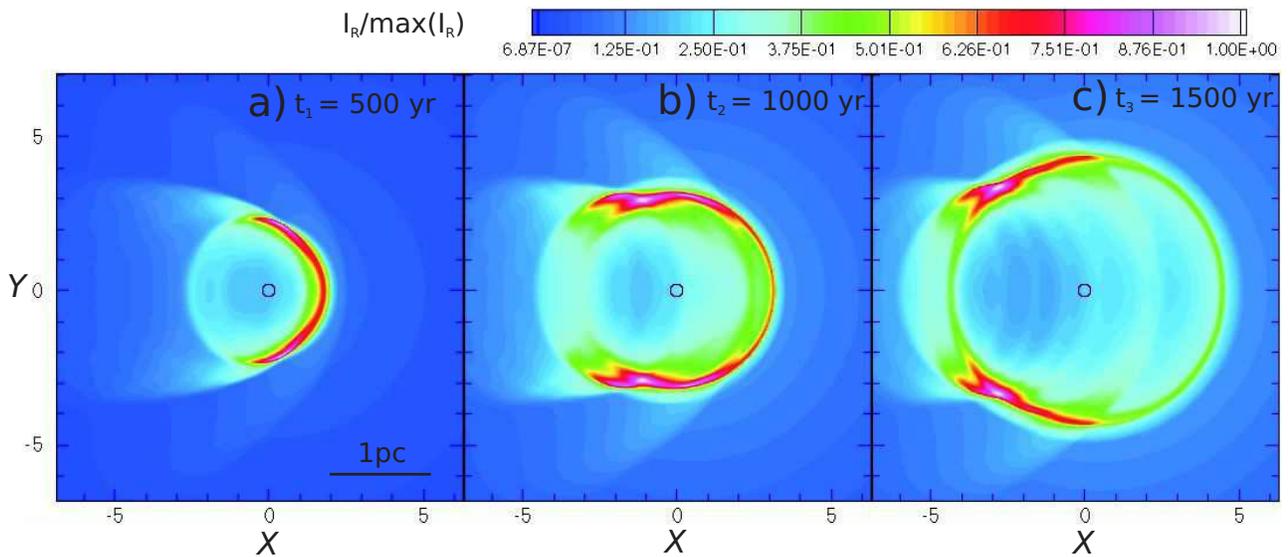}
\centering \caption{Synthetic maps of 22\,\micron \, dust emission (normalized to the maximum surface brightness of each panel) 
from the shell driven by the variable wind from the binary product at three times and for the inclination angle of 33\degr. 
The distance units on the $X$ and $Y$ axes are in 0.3 pc.} 
\label{fig:wint}
\end{figure*}
%%%%%%%%%%%%%%%%%%%%%%%%%%%%%%%%%%%%%%%%%%%%%%%%%%%%%%%%%%%%%%%%%%%%%%%%%%%

To model the interaction between the stellar wind and the ISM, we used the time-dependent Godunov scheme (Godunov et al. 
1979), which was applied to a curvilinear moving grid. The discontinuities in the flows were allocated by using the 
fitting-capturing technique (Myasnikov 1997; Izmodenov \& Alexashov 2015). For simplification of the numerical 
simulations, we used the shock capturing method for the inner (stellar wind) shock, while for the outer (bow) shock and 
the contact/tangential discontinuities the exact fitting was applied. When necessary, the outer boundary of the grid was also 
moving. 

To increase the resolution properties of the first order Godunov scheme a piecewise-linear distribution of the parameters 
inside the grid cells were constructed. To achieve the TVD (Total Variation Diminishing) property of the scheme the slope 
limiters with compression parameters were used (Chakravarthy \& Osher 1985). The resolution in time remained first order 
which is not critical because of the small time step. The applied numerical method conserves the number of cells between 
discontinuities (in shock layers) and, consequently, conserves the necessary resolution in thin shock layers.
This is important for the numerical accuracy, e.g., when the radiative losses are taken into account (in our simulations
we used the cooling function from Cowie, McKee \& Ostriker 1981). The simulations were performed in the reference frame of
the star on a 2D grid with $200\times200$ cells. For verification of numerical solutions the resolution was increased 
up to two times.

To produce synthetic maps of dust emission, we assumed that the dust grains are heated only due to absorption of the stellar 
radiation, which is re-emitted as a black body radiation. In this case, the temperature of the dust is related to the 
temperature and radius of the star as follows: $T_{\rm d}(r)=T_*(R_*/2r)^{(2/4+\beta)}$, where $r$ is the distance from the 
star, and the parameter $\beta$ ranges between 1 and 2, depending on the properties of the dust grains (see, e.g., Decin 
et al. 2012 and references therein). We also assumed that dust exist only in the ISM and the material of
the slow wind. We run simulations for two limiting values $\beta$ and found that the choice 
of a particular value affects mainly the magnitude of the surface brightness. For the sake of definiteness, in what follows 
we present the results obtained for $\beta=2$. Since the main goal of our simulations is to show that the general shape of 
the nebula cannot be explained in the framework of the bow shock scenario and because most of the input parameters (such as 
$n_{\rm ISM}$ and the wind parameters before and after the binary interaction) are unknown, we did not try to reproduce the absolute 
surface brightness of the nebula. Instead, in the synthetic dust emission maps we show the projected surface brightness 
normalized to the maximum value.   

In our modelling we assumed that the wind-blowing star is moving with a velocity of $160 \, \kms$ through the homogeneous 
ISM of temperature 8\,000\,K, and that there is no large-scale motions in the ISM. First, we modelled a bow shock produced by 
a star whose wind parameters ($\dot{M}$ and $v_\infty$) are identical to those of CPD$-$64\degr2731. We run this simulation to 
show that the nebula around CPD$-$64\degr2731 cannot be interpreted as a bow shock. For this modelling we used the number 
density of the ISM of $\approx0.4 \, {\rm cm}^{-3}$, which follows from equation\,(\ref{eqn:ram}) and the observed value of 
$R_0$ of $\approx0.9$ pc.

To verify our scenario for the origin of the nebula, we also run a second simulation in which the runaway star was 
assumed to be initially composed of two $\approx20 \, \msun$ components with identical $\dot{M}$ and $v_\infty$ of, respectively, 
$10^{-8} \, \myr$ and $1200 \, \kms$. The number density of the ISM was assumed to be equal to $0.04 \, {\rm cm}^{-3}$, as 
predicted for $z\approx0.5$\,kpc by the model of vertical density profile of the Galactic disk by Dickey \& Lockman (1990; 
see their fig.\,10). For these parameters the stand-off distance of the bow shock generated by the runaway system equals to 
$\approx0.2$\,pc. At some point, the wind velocity was reduced by a factor of 20 to mimic slow outflow from the 
binary product, and, for the sake of simplicity, it was assumed that the wind momentum rate $\dot{M}v_\infty$ 
did not change. Correspondingly, the shape of the interface between the unperturbed and shocked ISM did not change as well, 
while the circumstellar wind material became a factor of 400 denser. After several thousand years (the time-scale of thermal 
adjustment), the wind parameters of the binary product were set to be equal to those of CPD$-$64\degr2731, and the new, much 
faster, wind started to blow a bubble in the preceding slow wind material bounded in the upstream direction by the bow shock. 
In the simulations it was assumed that the stellar winds are spherically symmetric and that there is no instant (i.e. on a 
dynamical time-scale) mass loss (which is expected if CPD$-$64\degr2731 is a merger product). 

Figure\,\ref{fig:bow} shows synthetic maps of 22\,\micron \, dust emission from the steady-state bow shock produced by 
a runaway star with wind parameters and space velocity identical to those of CPD$-$64\degr2731. Panels a)--d) show maps
for four  inclination angles of the line of sight to the symmetry axis of the simulation ($Y=0$): 90\degr, 60\degr, 45\degr 
and 33\degr. The latter angle corresponds to the orientation of the space velocity vector of CPD$-$64\degr2731 derived from 
the proper motion and radial velocity measurements for this star. As expected, in all four projections the bow shock 
appears as an arc-like nebula with the associated star located close to the nose of the arc. This confirms that the 
nebula around CPD$-$64\degr2731 and the almost central location of this star in the nebula cannot be 
explained by the projection effect in the framework of the bow-shock model.  

In Fig.\,\ref{fig:wint} we show evolution of the wind bubble blown by the fast wind from the binary product after its
thermal adjustment. The panels plot synthetic maps of 22\,\micron \, emission for the inclination angle of 
33\degr. The shell around the bubble almost immediately appeared as an opened (arc-like) structure because of the presence 
of dense material ahead of the star (see panel a). At this time the nebula could be confused with a bow shock. After 
$\approx1\,000$ yr of evolution the bubble expanded in the forward direction beyond the region occupied by the material of 
the preceding slow wind and started to interact with the unperturbed ISM. At this time the shell adopts a horseshoe shape
(panel b). Later on (panel c) the shell became more rounded, while the maximum of its surface brightness has shifted 
downstream. At the same time, the star still remains close to the geometric centre of the bubble. At a later time (not shown 
in Fig.\,\ref{fig:wint}), after the star will completely overrun the region occupied by the material of the slow wind,
the bubble transformed into a bow shock identical to that shown in Fig.\,\ref{fig:bow}, except for a factor of $10^{1/2}$
larger stand-off distance due to a factor of 10 lower $n_{\rm ISM}$ adopted in the second simulation (cf. Meyer et al. 
2014). 

Although the model shell reproduces reasonably well the horseshoe shape of the nebula around CPD$-$64\degr2731, there is a 
notable distinction between them. Namely, the symmetry axis of the model nebula is parallel to the vector of stellar space 
velocity, while that of the observed nebula is misaligned by an angle of $\approx30$\degr (see Fig.\,\ref{fig:gal}). 
A possible explanation of this misalignment is that the star is moving at an angle to a density gradient in the 
local medium. The density inhomogeneity could be inherent to the local ISM or caused by non-spherically symmetric density 
distribution in the slow wind, which is expected to be concentrated towards the orbital plane of the binary system, whose 
orientation could be arbitrary with respect to the vector of stellar velocity. 

%%%%%%%%%%%%%%%%%%%%%%%%%%%%%%%%%%%%%%%%%%%%%%%%%%%%%%%%%%%%%%%%%%%%%%%%%%%
\begin{figure}
\includegraphics[width=8cm,angle=0]{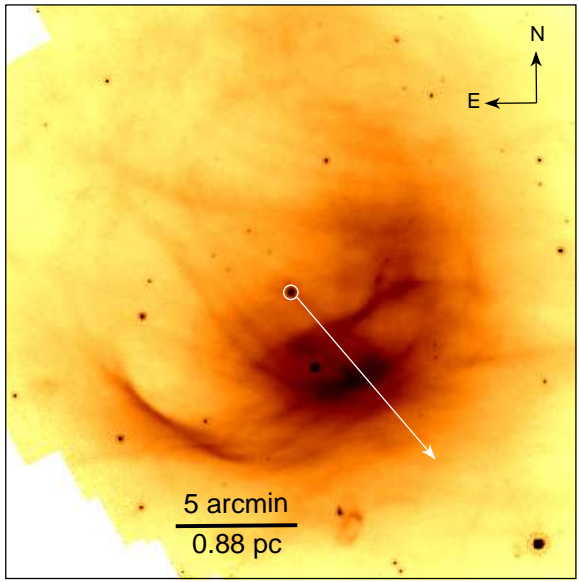}
\centering \caption{{\it Spitzer} 24\,\micron \, image of a semi-circular nebula around the runaway star $\lambda$\,Cep 
(marked by a circle). The arrow shows the direction of the peculiar transverse velocity of the star based on the 
{\it Gaia} DR2 parallax and proper motion measurements.} 
\label{fig:cep}
\end{figure}
%%%%%%%%%%%%%%%%%%%%%%%%%%%%%%%%%%%%%%%%%%%%%%%%%%%%%%%%%%%%%%%%%%%%%%%%%%%

\section{Concluding remarks}
\label{sec:con}

Our interpretation of the nebula around CPD$-$64\degr2731 as a transient shell driven by the variable wind from the 
product of binary interaction could also be applied to some other runaway stars whose nebulae do not look like bow shocks. 
For example, a semi-circular mid-infrared shell centred on the single (Gies 1987) runaway (Blaauw 1961) O6.5\,I(n)fp 
(Sota et al. 2011) star $\lambda$\,Cep (see Fig.\,\ref{fig:cep}) might be a young bubble created by the variable mass 
loss from the binary merger product. This possibility is in line with the suggestion by Walborn et al. (2010) that the Onfp 
phenomenon could be caused by binary mass transfer or stellar mergers. 

\begin{figure}
\includegraphics[width=8cm,angle=0]{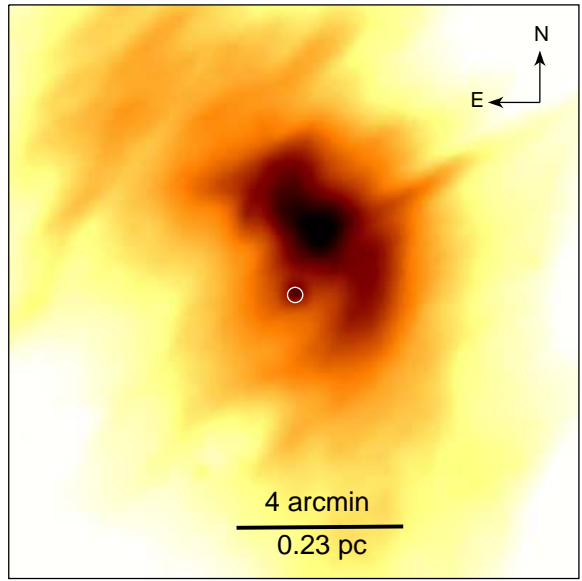}
\centering \caption{{\it WISE} 22\,\micron \, image of a nebula around the magnetic star $\tau$\,Sco (marked by a circle).} 
\label{fig:tau}
\end{figure}

The idea that merger of massive binaries play an important role in evolution and observational manifestations of OB stars 
attracts more and more attention during the last decade. For example, it was suggested that binary merger could be 
responsible for the origin of strong magnetic field observed in some massive stars (e.g. Ferrario et al. 2009; Langer 2012). 
In particular, Schneider et al. (2016) proposed that two magnetic stars, HR\,2949 (B3p IV) and  $\tau$\,Sco (B0.2\,V), are 
the rejuvenated merger products, and noted that such objects could be distinguished from genuine single stars by the presence 
of circumstellar material ejected before and/or during the merger process. Interestingly, $\tau$\,Sco is indeed surrounded by 
a curious infrared nebula (see Fig.\,\ref{fig:tau}). The unevolved status of this star implies that the nebula might have 
been formed because of binary merger, although one cannot also exclude the possibility that it represents a 
radiation-pressure-driven bow wave (cf. van Buren \& McCray 1988; Ochsendorf et al. 2014). 

%%%%%%%%%%%%%%%%%%%%%%%%%%%%%%%%%%%%%%%%%%%%%%%%%%%%%%%%%%%%%%%%%%%%%%%%%%%
%%%%%%%%%%%%%%%%%%%%%%%%%%%%%%%%%%%%%%%%%%%%%%%%%%%%%%%%%%%%%%%%%%%%%%%%%%%

Finally, we note that if CPD$-$64\degr2731 is a merger product of two stars then it could posses a strong magnetic field.
Thus, as noted above, spectropolarimetric observations of this star could be of high importance.

\section{Summary}
\label{sec:sum}

We have discovered an almost complete (horseshoe-shaped) mid-infrared shell around the high-velocity ($\approx160 \, 
\kms$) runaway star CPD$-$64\degr2731 with {\it WISE}. The shape of the shell and the nearly central location 
of CPD$-$64\degr2731 within it argue against the interpretation of the shell as a bow shock. Spectral analyses of 
CPD$-$64\degr2731, based on optical \'echelle spectroscopy with SALT, showed that we dial with a fast-rotating 
($v\sin i \approx300 \, \kms$) main-sequence O5.5 star with a factor of 6--8 
enhanced surface nitrogen abundance. The chemical abundances, high rotational velocity and large kinematic age of 
CPD$-$64\degr2731 exclude the possibility that this $\approx40 \, \msun$ star was ejected from the Galactic plane 
because of dissolution of a binary system following supernova explosion. Instead, we have proposed that CPD$-$64\degr2731 
was initially a compact binary system, which was dynamically ejected from a very massive ($\sim10^4 \, \msun$) star 
cluster about 6\,Myr ago and recently experienced strong binary interaction (either a mass transfer or merger) leading 
to spin-up and rejuvenation of this star. This interaction has also led to a strong and slow wind, either from the fast 
rotating accretion star or bloated merger product, lasting for a Kelvin-Helmholtz time-scale (about several thousand years). 
After the thermal adjustment the binary product established a fast wind which started to blow a bubble within the 
material of the preceding slow wind, producing a transient short-lived ($\sim1\,000$\,yr) shell reminiscent that around 
CPD$-$64\degr2731. This scenario was validated by means of 2D numerical modelling. We also have run a second simulation 
in which we modelled a bow shock produced by a runaway star whose wind parameters and space velocity were set to be 
identical to those of CPD$-$64\degr2731. With this simulation we confirmed that the nebula around CPD$-$64\degr2731 
cannot be explained by the projection effect in the framework of bow-shock model. We also have proposed that the 
semi-circular nebula centred on the runaway Onfp star $\lambda$\,Cep could be another case of a transient shell created 
because of changes in the mass-loss rate and wind velocity following the binary interaction (most likely a binary merger). 
Finally, we presented for the first time a {\it WISE} 22\,\micron \, image of a curious nebula around the main-sequence 
magnetic star $\tau$\,Sco, which is believed to be a rejuvenated binary merger product. This nebula could represent a 
circumstellar material ejected before and/or during the merger process or, alternatively, a radiation-pressure-driven 
bow wave.

\section{Acknowledgements}
We are grateful to I.D.\,Howarth (the referee) for useful comments and suggestions on the manuscript, and to J.\,Ma\'iz 
Apell\'aniz for useful discussion.
Spectroscopic observations, data reduction and spectral modelling were supported by the Russian Foundation for Basic 
Research grant 16-02-00148. Interpretation of the observational data and numerical simulations were supported by the 
Russian Science Foundation grant No. 14-12-01096. AYK acknowledges support from the National Research Foundation (NRF) 
of South Africa. This work was supported in part by M.V.\,Lomonosov Moscow State University Program of Development and
is based on observations collected with the Southern African Large Telescope (SALT), programmes 2017-1-MLT-005 and 
2018-1-MLT-008. This work has made use of data products from the Wide-field Infrared Survey Explorer, which is a joint 
project of the University of California, Los Angeles, and the Jet Propulsion Laboratory/California Institute of Technology, 
funded by the National Aeronautics and Space Administration, the SIMBAD data base and the VizieR catalogue access tool, 
both operated at CDS, Strasbourg, France, the WEBDA data base, operated at the Department of Theoretical Physics and 
Astrophysics of the Masaryk University, and data from the European Space Agency (ESA) mission 
{\it Gaia} (https://www.cosmos.esa.int/gaia), processed by the {\it Gaia} Data Processing and Analysis Consortium 
(DPAC, https://www.cosmos.esa.int/web/gaia/dpac/consortium). Funding for the DPAC has been provided by national 
institutions, in particular the institutions participating in the {\it Gaia} Multilateral Agreement.

\end{document}